\documentstyle[preprint,aps,floats]{revtex}

\tightenlines

\begin{document}

\renewcommand{\thefootnote}{\fnsymbol{footnote}}

\begin{flushright}
MPI-PhT 98-13, BI-TP 98/2\\
hep-ph/9802248
\end{flushright}

\vspace{0.5cm}

\centerline{{\large \bf 
Towards a physical expansion in perturbative gauge theories}}
\centerline{{\large \bf 
by using improved Baker-Gammel approximants}}

\vspace{1.cm}

\centerline{G.~Cveti\v c\footnote[1]{e-mail:
cvetic@doom.physik.uni-dortmund.de, cvetic@physik.uni-bielefeld.de}}

\centerline{{\it Max-Planck-Institut f\"ur Physik, P.O.B. 401212, 
80712 Munich, Germany, and}}

\centerline{{\it Department of Physics, Universit\"at Dortmund,
44221 Dortmund, Germany}}

\vspace{0.8cm}

\centerline{R.~K\"ogerler}

\centerline{{\it  
Department of Physics, Universit\"at Bielefeld,
33501 Bielefeld, Germany}}

\renewcommand{\thefootnote}{\arabic{footnote}}

\begin{abstract}

Applicability of the previously introduced method of modified 
diagonal Baker-Gammel approximants is extended to truncated 
perturbation series (TPS) of {\em any\/} order in gauge theories.
The approximants reproduce the TPS when expanded in power 
series of the gauge coupling parameter to the order of that 
TPS. The approximants have the favorable property of being 
exactly invariant under the change of the renormalization scale, 
and that property is arrived at by a generalization of the
method of the diagonal Pad\'e approximants. The renormalization 
scheme dependence is subsequently eliminated by a variant of
the method of the principle of minimal sensitivity (PMS). This is 
done by choosing the values of the renormalization-scheme-dependent
coefficients (${\beta}_2,\!{\beta}_3,\!\ldots$), which appear in 
the beta function of the gauge coupling parameter, in such a way 
that the diagonal Baker-Gammel approximants have zero values of 
partial derivatives with respect to these coefficients. The 
resulting approximants are then independent of the renormalization 
scale and of the renormalization scheme.\\ 
PACS number(s): 11.10.Hi, 11.80.Fv, 12.38.Bx, 12.38.Cy

\end{abstract}

\setcounter{equation}{0}

\section{Introduction}
Ordinary perturbation theory -- despite its intuitive physical
content as illustrated by the individual Feynman diagrams --
is plagued by several unpleasant features which obstruct a
strictly physical interpretation of various terms. The main
reason for this lies in the dependence on unphysical
structures like renormalization scale (RScl) and renormalization
scheme (RSch). This dependence represents a certain amount of
arbitrariness common to any finite order expression.
Considerable effort had been directed towards finding a 
pragmatic solution to a corresponding problem of defining an
appropriate RScl and RSch, respectively, for a given finite 
order expression.

Recently, a different method, 
involving modified diagonal Baker-Gammel
approximants (dBGA's), has been developed \cite{Cvetic}
for dealing with truncated perturbation series (TPS's) in gauge
theories. These approximants reproduce a TPS
to which they were applied,
when expanded to the order of that TPS. In addition, these
modified dBGA's were shown to be invariant
under the change of the renormalization scale (RScl) $q^2$,
i.e., when the evolution of the coupling parameter $a(q^2)$ 
in the TPS is determined by the full ${\beta}$-function
(to {\em any\/} chosen loop-order).
These dBGA's represent an improvement 
of the related method of diagonal Pad\'e approximants (dPA's).
The latter are RScl-invariant when $a(q^2)$ is
taken to evolve according to the one-loop beta function
(large-$|{\beta}_0|$ limit) \cite{Gardi}.
However, two remaining deficiencies of the dPA method
have not been eliminated in this way. Firstly, the
method appeared to be applicable only to the TPS's of the
observables $S$ with an odd number of nonleading terms.
Secondly, the approximants remained dependent on the
renormalization-scheme (RSch), i.e., on the values of the
RSch-dependent coefficients ${\beta}_j$ ($j\!\geq\!2$)
appearing in the beta function of the gauge coupling
parameter.

In this paper, we show that the first deficiency can be
cured easily, and that the method of modified dBGA's
(and also the method of dPA's) can be applied, in a
somewhat modified form, also to the TPS's with an even
number of nonleading terms. 
This is of high practical importance since nowadays --
while perturbation series of several QED observables
(e.g., anomalous magnetic moment of $e$ and $\mu$)
are available up to the third nonleading order  --
all interesting QCD quantities are only calculated to
at most second nonleading order, with next order corrections
not being expected for the near future.
We further show how to apply
the method when the leading term in the TPS of an observable
is proportional to $a^{\ell}(q^2)$ with $\ell\!>\!1$.

Also the second deficiency can be at least partially cured, and 
we can eliminate the RSch-dependence by applying a variant of the 
principle of minimal sensitivity (PMS) to the obtained dBGA's.
The method consists in finding such (RSch-dependent)
coefficients ${\beta}_j$ ($j\!=\!2,\!3,\!\ldots$)
for which the partial derivatives of the dBGA's
acquire the value zero. The obtained approximants then
possess RScl-invariance that was arrived at by a generalization
of the dPA method, and possess RSch-invariance\footnote{
This means that the approximants are the same, whatever
RSch-parameters ${\beta}_j$ ($j\!\geq\!2$) are used for
the TPS under consideration.}
that was arrived at by a variant of the PMS approach. In this
context, we mention that the original version of the PMS
\cite{Stevenson} was applied directly to the TPS's.
Since the method of the (d)PA's has proven to be remarkably
efficient for QCD observables \cite{PAQCD}, we expect
that the method presented here allows some room for optimism
as to its efficiency when compared with the usual PMS,
and with the method of effective charges (ECH)
\cite{Grunberg} that is related to the usual PMS.\footnote{
The usual PMS and the ECH methods consist in defining 
in a pragmatic way an appropriate RScl and RSch
for a given finite order expression (TPS).}

\section{Extending the method to any truncated series}
An observable $S$ in a gauge theory 
can in general be redefined so that it has the following
form as a formal perturbation series:
\begin{equation}
S^{(\ell)} \equiv a^{\ell}(q^2) f(q^2) = a^{\ell}(q^2) \left[
1 + r_1(q^2) a(q^2) + \cdots 
+ r_n(q^2) a^n(q^2) + \cdots \right] \quad (\ell \geq 1) \ ,
\label{Sp}
\end{equation}
where $q^2$ is the chosen renormalization scale (RScl).
The observable is, of course, independent of the
RScl $q^2$.
The series is known in practice only up to an
$n$'th order in nonleading terms
\begin{equation}
S^{(\ell)}_n(q^2) \equiv a^{\ell}(q^2) f^{(n)}(q^2) 
= a^{\ell}(q^2) \left[
1 + r_1(q^2) a(q^2) + r_2(q^2) a^2(q^2) + \cdots 
+ r_n(q^2) a^n(q^2) \right] \ .
\label{Spn}
\end{equation}
This TPS explicitly depends on the RScl $q^2$, due
to truncation. In most practical cases 
we have $\ell\!=\!1$, and at present
we have at most $n\!=\!3$ in QED and $n\!=\!2$ in QCD.
In our previous work \cite{Cvetic} we constructed
modified dBGA's with kernel $a(p^2)/a(q^2)$ for TPS
(\ref{Spn}) when ${\ell}\!=\!1$ and $n\!=\!2M\!-\!1$
($M\!=\!1,\!2,\ldots$). The latter condition ($n$ odd)
originated from the need to employ {\em diagonal\/}
Pad\'e approximants (dPA's) in the algorithm, since only
such PA's are RScl-independent in the large-$|{\beta}_0|$
limit.

Now we show how to modify the
mentioned method so that it is applicable also to
the cases of $n\!=\!2M\!-\!2$ ($M\!=\!2,\!3,\!\ldots$)
and ${\ell}\!=\!1$. For this purpose we consider instead of
\footnote{
We omit superscript $(\ell)$ when $\ell\!=\!1$.} 
the observable $S$ its square, i.e., we construct the observable
\begin{equation}
{\tilde S} \equiv S * S \equiv a(q^2) F(q^2) 
= a(q^2) \left[ 0 +
1 \cdot a(q^2) + {R}_2(q^2) a^2(q^2) + 
{R}_3(q^2) a^3(q^2) + \ldots \right] \ ,
\label{calS}
\end{equation}
where coefficients ${R}_j(q^2)$ are in general
again RScl-dependent and are related to the expansion
coefficients $r_i(q^2)$'s in the following way:
\begin{eqnarray}
{R}_1 &=& 1 \ , \quad
{R}_2 = 2 r_1 \ , \quad
{R}_3 = 2 r_2 + r_1^2 \ , \quad \ldots
\nonumber\\
{R}_{2M-1} &=& 2 r_{2M-2} + 2 r_{2M-3} r_1 +
2 r_{2M-4} r_2 + \cdots + 2 r_M r_{M-2} + r^2_{M-1} \ ,
\qquad \ldots
\label{calrr}
\end{eqnarray}
When the observable $S$ is available up to an even nonleading order,
say $S_{2M-2}$ is available
[i.e., $r_1(q^2),\!\ldots,\!r_{2M-2}(q^2)$ are known],
then according to (\ref{calrr}) all coefficients of ${\tilde S}$
up to $R_{2M-1}(q^2)$ are known and consequently we
know ${\tilde S}_{2M-1}(q^2)$ 
\begin{eqnarray}
\lefteqn{
{\tilde S}_{2M-1}(q^2) \equiv a(q^2) {F}^{(2M-1)}(q^2) } 
\nonumber\\
&=& a(q^2) \left[ 0 +
1\!\cdot\!a(q^2) + {R}_2(q^2) a^2(q^2) +
{R}_3(q^2) a^3(q^2) + \ldots +{R}_{2M-1}(q^2) a^{2M-1}(q^2) \right] \ .
\label{calS2M-1}
\end{eqnarray}
This TPS has formally ${\ell}\!=\!1$ and the number of
nonleading terms is formally odd ($2M-1$). Therefore,
we now just repeat the procedure of constructing
dBGA's described in \cite{Cvetic} for the TPS's with
odd number of nonleading terms -- the only difference being now
that the formal leading term is zero instead of one, 
and ${R}_1(q^2)\!\equiv\!1$ is RScl-independent.
First we recall that the gauge coupling parameter 
$a(p^2)\!\equiv\!{\alpha}(p^2)/\pi$ evolves according to
the perturbative renormalization group equation (RGE)
\begin{equation}
\frac{d a(p^2)}{d \ln p^2} = 
- \sum_{j=0}^{\infty} {\beta}_j a^{j+2}(p^2) \ ,
\label{aRGE}
\end{equation}
where ${\beta}_j$ are 
constants if particle threshold effects
are ignored. We now introduce for two different
scales $p^2$ and $q^2$ the ratio 
\begin{equation}
k( a_q,u ) \equiv \frac{ a(p^2) }{ a(q^2) }
\qquad \mbox{where: } \quad a_q=a(q^2) \ , \ u=\ln(p^2/q^2) . \ 
\label{kdef}
\end{equation}
Formal expansion of this function in powers of $u$ contains
coefficients $k_j\!\sim\!a^j(q^2)$
\begin{equation}
k( a_q,u ) = 1 + \sum_{j=1}^{\infty} u^j k_j(a_q) \ ,
\  \mbox{with: } k_j(a_q) = \frac{1}{j!} 
\frac{\partial^j}{\partial u^j} k(a_q, u) {\Big |}_{u=0} 
= \frac{1}{j! a(q^2)} 
\frac{d^j a(p^2) }{d (\ln p^2)^j}{\Bigg |}_{p^2=q^2}\ .
\label{kTaylor}
\end{equation}
These coefficients can be determined from RGE (\ref{aRGE})
as a series in powers of $a(q^2)$
\begin{equation}
k_j(a_q) = (-1)^j \beta_0^j a^j(q^2) + 
{\cal O}\left( a^{j+1}(q^2) \right) \ , \quad
k_0(a_q) = 1 \ ,
\label{kjsapprox}
\end{equation}
We rearrange the series for 
${\tilde S}/a(q^2)\!\equiv\!F(q^2)$ of (\ref{calS})
into the series in powers of $k_j(a(q^2))$
\begin{equation}
{\tilde S} \equiv S * S \equiv a(q^2) {F}(q^2) = 
a(q^2) \left[ 0 + \sum_{j=1}^{\infty} {F}_j(q^2) 
k_j \left( a(q^2) \right) \right] \ ,
\label{calSinkj}
\end{equation} 
and define the corresponding formal series obtained from the
above by replacing the coefficients 
$k_j(a(q^2))$ by their large-$|{\beta}_0|$
limits $(-{\beta}_0 a(q^2))$ [cf.~(\ref{kjsapprox})]
\begin{equation}
a(q^2) {\cal F}(q^2) \equiv a(q^2) \left[
0 + \sum_{j=1}^{\infty} F_j(q^2) 
(- 1)^j {\beta}_0^j a^j(q^2) \right] \ .
\label{F1}
\end{equation} 
Now we construct for this expression the diagonal Pad\'e approximant
(dPA) of order $2M$, with\footnote{
Note that this dPA, by construction, has a polynomial of 
order $M$ [in $a(q^2)$] in the nominator and a polynomial of 
the same order in the denominator.} 
argument $a(q^2)$
\begin{equation}
a(q^2) [ M-1 / M ]_{ {\cal F} }(q^2) =
a(q^2) \left[ 0 + \sum_{m=1}^{M-1} 
{\tilde a}_m(q^2) a^m(q^2) \right]
\left[ 1 + \sum_{n=1}^{M} {\tilde b}_n(q^2) a^n(q^2) \right]^{-1} \ .
\label{PAF11}
\end{equation}
By construction, this dPA satisfies the relation
\begin{equation}
a(q^2) {\cal F}(q^2) =
a(q^2) [ M-1 / M ]_{ {\cal F} }(q^2) + 
{\cal O} \left( a^{2M+1}(q^2) \right) \ .
\label{PAF12}
\end{equation}
The above dPA depends only on the first $(2M\!-\!1)$ coefficients
$F_j(q^2)$ ($j\!=\!1,\!\ldots,\!2M\!-\!1$), as follows from the
standard PA relation (\ref{PAF12}). The 
latter coefficients are uniquely determined by the coefficients
$R_1(q^2)\!=\!1,\!\ldots,\!R_{2M-1}(q^2)$ of the series
(\ref{calS2M-1}) via relations
(\ref{kjsapprox}), and these coefficients are determined
uniquely by the initial TPS 
$S^{(1)}_{2M-2}\!\equiv\!S_{2M-2}$ of (\ref{Spn}),
i.e., by the first $2M\!-\!2$ nonleading terms of
the observable $S^{(1)}\!\equiv\!S$ 
[$r_1(q^2),\!\ldots,\!r_{2M-2}(q^2)$] as seen from
relations (\ref{calrr}). Therefore, knowing
$S_{2M-2}$ ($M\!\geq\!2$), we can uniquely construct the above dPA
(\ref{PAF11})-(\ref{PAF12}).

If we don't have an exceptional situation when the
denominator in the dPA (\ref{PAF11}) has multiple zeros,
we can uniquely decompose this dPA into a sum of
simple fractions
\begin{equation}
a(q^2) [ M-1 / M ]_{ {\cal F} }(q^2) =
a(q^2) \sum_{i=1}^M \frac{ {\tilde \alpha}_i }
{ \left[ 1 + {\tilde u}_i(q^2) {\beta}_0 a(q^2) \right] } 
= \sum_{i=1}^M {\tilde \alpha}_i  \frac{ a(q^2) }
{ \left[ 1 + {\tilde u}_i(q^2) {\beta}_0 a(q^2) \right] } \ .
\label{PAF1decomp}
\end{equation}
Here, $[-{\tilde u}_i(q^2) {\beta}_0]^{-1}$ are the $M$ zeros of the
denominator of the dPA (\ref{PAF11}). We note that the above sum is
a weighed sum of one-loop-evolved gauge coupling parameters
$a(p_i^2)$, where the generally complex scales $p_i^2$ are
determined by relation ${\tilde u}_i(q^2)\!=\!\ln(p_i^2/q^2)$,
and the one-loop evolution (i.e., in the large-$|{\beta}_0|$
limit) is performed from the RScl $q^2$ to $p_i^2$.
The dBGA approximant is then obtained by replacing in the
above sum the one-loop-evolved gauge coupling parameters
by those which are evolved according to the full RGE
(\ref{aRGE}) from the RScl $q^2$ to $p_i^2$
\begin{equation}
a(q^2) G_{ F }^{ [M-1/M] }(q^2) \equiv 
\sum_{i=1}^M {\tilde \alpha}_i a(p_i^2) \ , \qquad
\mbox{where: } \quad 
p_i^2 = q^2 \exp \left[ {\tilde u}_i (q^2) \right] \ .
\label{dBGAres}
\end{equation}
We emphasize that this dBGA is an approximant for the
squared observable ${\tilde S}\!\equiv\!\!S*S$.
As in Refs.~\cite{Cvetic}, we can show explicitly that
the above approximant fulfills two properties:
\begin{enumerate}
\item
It has the same formal accuracy as the TPS 
${\tilde S}_{2M-1}(q^2)$ of (\ref{calS2M-1}):
\begin{equation}
{\tilde S} = a(q^2) G_{ F }^{ [M-1/M] }(q^2) + 
{\cal O} \left( a^{2M+1}(q^2) \right) \ .
\label{Theor1}
\end{equation} 
\item
It is invariant under the change of the renormalization
scale (RScl) $q^2$, where the evolution from one to another
RScl is performed according to
RGE (\ref{aRGE}) with any chosen loop precision.\footnote{
We should keep in mind, however, that only a limited number 
of the perturbative coefficients ${\beta}_j$ 
(${\beta}_0,\!\ldots,\!{\beta}_3$), appearing on the
right of RGE (\ref{aRGE}) have been calculated
and are consequently known in QCD
(cf.~\cite{QCDbeta}, in ${\overline {\mbox{MS}}}$ scheme) and in 
QED (cf.~\cite{QEDbeta}, in ${\overline {\mbox{MS}}}$, MOM and
in on-shell schemes). Hence, in practice, RGE (\ref{aRGE})
has to be truncated at the four-loop level.}
Incidentally, the weights ${\tilde \alpha}_i$ and the
scales $p_i^2\!=\!q^2 \exp[ {\tilde u}_i(q^2) ]$
are separately independent of the chosen RScl $q^2$.
\end{enumerate}
The proof of these two statements can be taken over word by word
from Ref.~\cite{Cvetic} (first entry) where the approximated TPS
was $S_{2M-1}$ and not ${\tilde S}_{2M-1}$. The only formal
difference in the dBGA (\ref{dBGAres}) is that the
sum of the ${\tilde \alpha}_i$ parameters is now zero and
not 1. This is due to the fact that the series for ${\tilde S}$,
in the form written in (\ref{calS}), has formally the
leading term [$\propto\!a(q^2)$] equal to zero,
while this term in $S$ in Ref.~\cite{Cvetic} has coefficient $1$.

The last step is to take simply the square root of this
expression, and this gives us an effective modified dBGA approximant
to the TPS $S^{(1)}_{2M-2}\!\equiv\!S_{2M-2}$, 
up to and including terms $\sim\!a^{2M-1}$, and this approximant is, 
of course, again RScl-invariant
\begin{equation}
S = \left[ a(q^2) G_{ F }^{ [M-1/M] }(q^2) \right]^{1/2} + 
{\cal O} \left( a^{2M}(q^2) \right) \ .
\label{sqdBGA}
\end{equation}
We mention that the parameters ${\tilde \alpha}_i$ and
${\tilde u}_i$ can be in general nonreal (complex). Then
the modified dBGA (\ref{dBGAres}) could possibly be nonreal, too. 
However, in this case we would just take 
under the square root in (\ref{sqdBGA}) the
real part of the dBGA (\ref{dBGAres}) -- it is also RScl-invariant,
it also satisfies relation (\ref{Theor1}) because observable
${\tilde S}$ is real, and it is positive because 
${\tilde S}\!\equiv\!S*S$ is positive. Below we will see,
however, that in the practically interesting case
of $n\!\equiv\!2M\!-\!2\!=2$ ($M\!=\!2$) the approximant
under the square root in (\ref{sqdBGA}) is always real
(and positive).

Let us illustrate this somewhat abstract deliberations
in the specific case of $M\!=\!2$ ($S_{2M-2}\!=\!S_2$)
which is at present of particular interest for several
QCD observables. The starting point is then the knowledge of
the two coefficients $r_1(q^2)$ and $r_2(q^2)$ of the
TPS $S^{(1)}_2\!\equiv\!S_2$ of Eq.~(\ref{Spn}), at a given RScl
$q^2$ and in a specific renormalization scheme. We then
also know the coefficients $R_1\!=\!1$, $R_2(q^2)$ and
$R_3(q^2)$ of the TPS (\ref{calS2M-1}) via relations
(\ref{calrr}), i.e., we know ${\tilde S}_3$ there.
Explicit form of relations (\ref{kjsapprox}), obtained
from RGE (\ref{aRGE}), yields 
\begin{eqnarray}
k_1(a) & = & - {\beta}_0 a - {\beta}_1 a^2 - {\beta}_2 a^3 
- \ldots \ ,
\label{k1}
\\
k_2(a) &=& + {\beta}_0^2 a^2 + (5/2) {\beta}_0 {\beta}_1 a^3 +
\ldots \ , \qquad k_3(a) = - {\beta}_0^3 a^3 - \ldots \ .
\label{k2k3}
\end{eqnarray}
We use the short-hand notation $a\!\equiv\!a(q^2)$. Inverting
these relations gives us expressions for $a,\!a^2$ and $a^3$
in terms of $k_1,\!k_2$ and $k_3$, when terms of order
$k_4\!\sim\!a^4$ are neglected. Inserting the latter
relations into the series (\ref{calS2M-1}), 
we obtain the first three coefficients of the rearranged
series (\ref{calSinkj})
\begin{eqnarray}
F_1(q^2) &=& - \frac{1}{ {\beta}_0 } \ ,
\qquad
F_2(q^2) = - \frac{ {\beta}_1 }{ {\beta}_0^3 } 
+ \frac{1}{ {\beta}_0^2 } R_2(q^2) \ ,
\label{f11f12}
\\
F_3(q^2) &=& \left( - \frac{ 5 {\beta}_1^2 }{ 2 {\beta}_0^5 }
+ \frac{ {\beta}_2 }{ {\beta}_0^4 } \right)  
+ \frac{ 5 {\beta}_1 }{ 2 {\beta}_0^4 } R_2(q^2)
- \frac{1}{ {\beta}_0^3 } R_3(q^2) \ .
\label{f13}
\end{eqnarray}
On the basis of these three coefficients, we can write down the
truncated series for $a(q^2) {\cal F}(q^2)$ of (\ref{F1}),
and can subsequently construct the dPA
$a(q^2) [1/2]_{ {\cal F} }(q^2)$ of Eq.~(\ref{PAF12}) 
in the decomposed form
(\ref{PAF1decomp}). We then obtain expressions for
parameters ${\tilde u}_i(q^2)$ and ${\tilde \alpha}_i$ 
($i\!=\!1,\!2$) 
\begin{equation}
{ {\tilde u}_2 \choose {\tilde u}_1 } =
\frac{1}{2 {\beta}_0 } \left[ {\tilde b}_1 
\pm \sqrt{ {\tilde b}_1^2
- 4 {\tilde b}_2 } \right] \ ,
\quad
{\tilde \alpha}_1  =  \frac{ {\beta}_0 {\tilde u}_1 {\tilde u}_2 }
{ {\tilde b}_2 ( {\tilde u}_2 - {\tilde u}_1 ) } = 
\frac{1}{ \sqrt{ {\tilde b}_1^2 - 4 {\tilde b}_2 } } = 
- {\tilde \alpha}_2
\ ,
\label{param}
\end{equation}
where ${\tilde b}_1$ and ${\tilde b}_2$ are the two coefficients
in the denominator of the dPA (\ref{PAF11}) (for the case
$2M\!-\!2\!=\!2$)
\begin{equation}
{\tilde b}_1 = \frac{ {\beta}_1 }{ {\beta}_0 } - 2 r_1 \ , \
{\tilde b}_2 =  \left( - \frac{ 3 {\beta}_1^2 }{ 2 {\beta}_0^2 }
+ \frac{ {\beta}_2 }{ {\beta}_0 } \right) +
\frac{ {\beta}_1 }{ {\beta}_0 } r_1 + 3 r_1^2 - 2 r_2 \ .
\label{paramnot}
\end{equation}
We note that there are basically two cases to be distinguished
when $2M\!-\!2\!=\!2$:
\begin{enumerate}
\item
The discriminant in (\ref{param}) is nonnegative; then all the
above parameters ${\tilde u}_i$, ${\tilde \alpha}_i$
and $p_i^2\!=\!q^2 \exp( {\tilde u}_i )$ are real, and the
dBGA (\ref{dBGAres}) is a real (and positive) number.
\item
The discriminant in (\ref{param}) is negative. Then 
${\tilde u}_2\!=\!{\tilde u}_1^{\ast}$, $p_2^2\!=\!(p_1^2)^{\ast}$,
$a(p_2^2)\!=\!a(p_1^2)^{\ast}$; and ${\tilde \alpha}_1$ is
imaginary. Therefore, by (\ref{param}),
the dBGA (\ref{dBGAres}) is a product of two imaginary numbers 
${\tilde \alpha}_1$ and $a(p_1^2)\!-\!a(p_1^2)^{\ast}$,
and hence is real (and positive).
\end{enumerate}
It is interesting to establish directly RScl-invariance of
the dBGA's (\ref{dBGAres}) and (\ref{sqdBGA}) for the case
$2M\!-\!2\!=\!2$. Based on formulas (\ref{param})-(\ref{paramnot}),
it is straightforward to show that the parameters
${\tilde \alpha}_i$ and $\ln (p_i^2/ {\tilde \Lambda}^2 )$
are functions of only the RScl-invariants
${\rho}_1\!\equiv\!{\tau}\!-\!r_1(q^2)$. Here,
${\tau}\!\equiv\!{\beta}_0 \ln( q^2/{\tilde \Lambda}^2 )$ was
introduced in \cite{Stevenson} and is just a dimensionless
version\footnote{
We consider the renormalization schemes in which the
RScl-parameter $q^2$ has always the same meaning, i.e., 
the energy parameter ${\tilde \Lambda}$ is the same
in all considered schemes. See also discussion later on.}
of the RScl parameter $q^2$. However, the RSch-invariance,
as expected, is not fulfilled in the 
$2M\!-\!2\!=\!2$ case, due to appearance
of the RSch-dependent coefficient ${\beta}_2$ in the mentioned
parameters.

As mentioned earlier, normalized observables $S^{(\ell)}$
appear sometimes in a modified form (\ref{Sp}) with $\ell$ 
larger than $1$. In such a case, we can simply introduce
another observable $S\!=\!(S^{(\ell)})^{1/\ell}$ which is
then represented as a power series with effectively 
${\ell}\!\mapsto\!1$, as application of the simple Taylor
expansion formula for $(1+x)^{1/\ell}$ in powers of $x$ shows.

We should keep in mind that Ref.~\cite{Cvetic} 
presented a dBGA algorithm applicable directly
to truncated perturbation series (\ref{Spn}) with
${\ell}\!=\!1$ and $n$ being an {\em odd\/} positive integer.
Here we extended application of this algorithm to
the case of (\ref{Spn}) with $n$ being an {\em even\/}
positive integer ($n\!\equiv\!2M\!-\!2$) and ${\ell}\!=\!1$.
Therefore, combining the results of 
\cite{Cvetic} and those of the present paper, 
keeping in mind also the mentioned trick of
reducing the ${\ell}\!\not=\!1$ to the ${\ell}\!=\!1$ case,
we see that we can apply the method of the modified dBGA's 
to {\em any\/} available truncated perturbation series of 
any normalized observable (\ref{Sp}). The modified dBGA's
are those with the kernel $k(a_q,u)$ defined via (\ref{kTaylor}),
they reproduce the available TPS up to the order to which
that TPS is known, and they are globally RScl--invariant.

\section{Improving the approximants by a PMS variant}
The described method of modified dBGA's yields
RScl-invariant approximants, i.e., the approximants
are independent of the choice of the RScl $q^2$ in the
TPS under consideration. However, at this stage,
the method does not address the question of the RSch-dependence. 
As already argued by Stevenson \cite{Stevenson}, 
the coefficients ${\beta}_j$ ($j\!=\!2,\!3,\ldots$) which 
show up in RGE (\ref{aRGE}) are not 
just RSch-dependent,\footnote{
The coefficients ${\beta}_0$ and ${\beta}_1$ are RSch-independent.}
but their values in turn characterize an RSch. The sets of
values $\{ {\beta}_j; j\!=\!2,\!3,\ldots \}$ thus represent
a convenient parametrization of RSch's.
In fact, the obtained dBGA's are in general
dependent on the RSch parameters ${\beta}_j$ ($j\!\geq\!2$)
appearing in RGE (\ref{aRGE}).
However, in principle, we can achieve RSch-independence 
by requiring that the corresponding partial derivatives
of the dBGA's be zero
\begin{equation}
\frac{ {\partial} }{ {\partial \beta}_j } 
\left( a G_f^{[M-1/M]} \right) = 0 \quad (j=2,3,\ldots) \ .
\label{PMSeq}
\end{equation}
We should recall that the RSch-invariance of the
approximants means that they are independent of the
particular choice of the RSch-parameters ${\beta}_j$ 
($j\!\geq\!2$) made in the original TPS under consideration.
Looking at conditions (\ref{PMSeq}), we should keep in mind that
the gauge coupling parameter $a\!\equiv\!{\alpha}/{\pi}$
and the coefficients $r_j$ in (\ref{Sp})
depend not just on the RScl $q^2$ [i.e., on 
${\tau}\!=\!{\beta}_0 \ln ( q^2/{\tilde \Lambda}^2 )$], but also
on the RSch-parameters ${\beta}_2,\!{\beta}_3,\!\ldots$,
and that the RScl-independent parameters ${\tilde \alpha}_i$
and $p_i^2$ appearing in the dBGA's (\ref{dBGAres}) also
depend on these RSch-parameters. As a matter of fact,
Eqs.~(\ref{PMSeq}) represent just the principle of minimal
sensitivity (PMS) introduced by Stevenson \cite{Stevenson}.
The difference now is that the PMS is applied to the dBGA's
which are already RScl-invariant, while the usual
PMS \cite{Stevenson} is applied to the TPS's 
which, at the outset, are by definition not just RSch-dependent, 
but also RScl-dependent. In a way,
we repeat the PMS approach with 
a different, presumably more favorable,
set of functions.

In general, the dBGA applied to an available TPS of an
appropriately redefined observable $S$ (with effective 
${\ell}\!=\!1$)
is written in the form (\ref{dBGAres}). Therefore, the PMS
Eqs.~(\ref{PMSeq}) can be rewritten as
\begin{equation}
\frac{ {\partial} }{ {\partial c}_j } 
\left( a G_f^{[M-1/M]} \right) =
\sum_{i=1}^{M} \frac{ {\partial} 
{\tilde \alpha}_i }{ {\partial} c_j } a(p_i^2) +
  \sum_{i=1}^{M} {\tilde \alpha}_i \left[
\frac{ {\partial} a(p_i^2) }{ {\partial} (p_i^2) }
\frac{ {\partial} (p_i^2) }{ {\partial} c_j } +
\frac{ {\partial} a(p_i^2) }{ {\partial} c_j }{\Bigg |}_{p_i^2} 
\right] = 0 \ ,
\label{PMSeq2}
\end{equation}
where the RSch-parameters $c_j$ are defined in the conventional way
\begin{equation}
c_j \equiv \frac{ {\beta}_j }{ {\beta}_0 } \ \
(j \geq 2) \ ,  \qquad 
\frac{ {\partial} }{ {\partial} c_j } =
{\beta}_0 \frac{ {\partial} }{ {\partial} {\beta}_j } \ .
\label{cjs}
\end{equation}
The derivatives in (\ref{PMSeq2}) are partial in the sense that
all other $c_k$'s ($k\!\not=\!j$) are kept constant, as well
as the RScl $q^2$ (although the entire expression is
independent of $q^2$).

In the case of the TPS (\ref{Spn}) with $n\!=\!1$, i.e.,\footnote{
From now on, ${\ell}\!=\!1$ will be assumed.} when only
one term beyond the leading term is known, the dBGA
depends explicitly only on the RScl $q^2$ (or: ${\tau}$) 
and not on $c_j$'s (cf.~\cite{Cvetic}). Therefore, the PMS
approach cannot be applied in this case. As a matter of fact,
for the $n\!=\!1$ case the dBGA approach gives the same result
as the effective charge method (ECH) \cite{Grunberg}.
The PMS improvement of the dBGA method comes into effect
when higher order terms ($n\!=\!2,\!3,\!\ldots$) are available.
For example, we can directly apply the PMS improvement
in the case $n\!=\!2M\!-\!1\!=\!3$ ($M\!=\!2$),
by using explicit formulas provided in Ref.~\cite{Cvetic}
for this case. In this case, we have coefficients $f_j$
($j\!=\!1,\!2,\!3$) instead of $F_j$'s of (\ref{F1}),
and these $f_j$'s are explicitly given in terms of
$r_j$'s and ${\beta}_j$'s in Ref.~\cite{Cvetic}.
Now, using the RScl- and RSch-invariant quantities
${\rho}_1,\!{\rho}_2,\!{\rho}_3$ introduced by Stevenson
\cite{Stevenson} and derived from the available coefficients
$r_1,\!r_2,\!r_3$ of the TPS (\ref{Sp}) (with $\ell\!=\!1$),
we have
\begin{eqnarray}
r_1 & = & {\tau} - {\rho}_1 \ , \
r_2 = {\rho}_2 + ({\tau}- {\rho}_1)^2 + c ( {\tau} - {\rho}_1 )
- c_2 \ ,
\label{r1r2}
\\
r_3 &=& {\rho}_3 - \frac{1}{2}c_3 + r_1 \left[ 2 r_2 - r_1^2
+ \frac{c}{2} r_1 + \frac{c^2}{4} + {\rho}_2 \right] \ ,
\label{r3}
\end{eqnarray}
where $c\!\equiv\!{\beta}_1/{\beta}_0$ 
is also RScl- and RSch-invariant,
and ${\tau}\!\equiv\!{\beta}_0 \ln( q^2/{\tilde \Lambda}^2 )$ 
is a dimensionless form of the RScl parameter $q^2$. 
This means that, when having the TPS $S_3$ 
(i.e., $r_1,\!r_2,\!r_3$) available in a certain RSch
and at a given RScl, we have $S_3$ available in {\em any\/}
RSch and at any RScl. From (\ref{r1r2})-(\ref{r3})
and from explicit expressions for $f_j$'s we obtain
\begin{eqnarray}
\frac{ {\partial} f_1 }{ {\partial} c_j } & = & 0 
\quad (j = 2,3, \ldots) \ ; \qquad \quad
\frac{ {\partial} f_2 }{ {\partial} c_2 } =
- \frac{1}{ {\beta}_0^2 } \ , 
\quad \frac{ {\partial} f_2 }{ {\partial} c_j } = 0
\quad (j = 3, \ldots ) \ ;
\label{parf1}
\\
\frac{ {\partial} f_3 }{ {\partial} c_2 } &=& 
\frac{3 r_1}{ {\beta}_0^3 } - \frac{ 5 {\beta}_1 }
{2 {\beta}_0^4 } \ ,
\quad \frac{ {\partial} f_3 }{ {\partial} c_3 } = 
\frac{1}{ 2 {\beta}_0^3 } \ , \quad
\quad \frac{ {\partial} f_3 }{ {\partial} c_j } = 0
\quad (j = 4, \ldots ) \ .
\label{parf3}
\end{eqnarray}
Using these identities, as well as the explicit expressions for the
dBGA parameters ${\tilde u}_i\!=\!\ln(p_i^2/q^2)$
and ${\tilde \alpha}_i$ ($i\!=\!1,\!2$) of Ref.~\cite{Cvetic},
we obtain explicit expressions for
the partial derivatives of these parameters with respect to
$c_j$'s, in terms of the original TPS coefficients $r_j$
(or equivalently: $f_j$)
\begin{eqnarray}
\frac{ \partial {\tilde u}_{2,1} }{ \partial c_2 } & = &
\frac{1}{p^2_{2,1}}\frac{ \partial {p}^2_{2,1} }{ \partial c_2 } =
\frac{1}{ (f_2-f_1^2) }
{\Bigg \{} \left( - \frac{1}{{\beta}_0^2} f_1 -
\frac{5 {\beta}_1}{4 {\beta}_0^4} \right)
\nonumber\\
&& \mp \frac{3}{ {\beta}_0^2 } \frac{1}{ \sqrt{{\rm det}} }
\left[ (f_2-f_1^2)^2 + \frac{5 \beta_1}{ 12 \beta_0^2 }
( f_3 - 3 f_1 f_2 + 2 f_1^3 ) \right] {\Bigg \}} 
 + \frac{ {\tilde u}_{2,1} }{ {\beta}_0^2 (f_2-f_1^2) } \ ,
\label{duc2}
\\
\frac{ \partial {\tilde u}_{2,1} }{ \partial c_3 } & = &
\frac{1}{p^2_{2,1}}\frac{ \partial {p}^2_{2,1} }{ \partial c_3 } =
= \frac{1}{4 {\beta}_0^3 (f_2-f_1^2) }
{\Bigg \{} 1 \pm \left[ f_3 - 3 f_1 f_2 + 2 f_1^3) \right]
\frac{1}{ \sqrt{ {\rm det} } } {\Bigg \}}  \ ,
\label{duc3}
\\
\frac{ {\partial} {\tilde \alpha}_1 }{ {\partial} c_2 } & = &
- \frac{ {\partial} {\tilde \alpha}_2 }{ {\partial} c_2 } =
\frac{3}{ {\beta}_0^2 ( {\rm det} )^{3/2} }
(f_2-f_1^2)^2 \left[ (f_3 - 3 f_1 f_2 + 2 f_1^3 ) 
- \frac{ 5 \beta_1 }{ 3 \beta_0^2 } (f_2 - f_1^2) \right] \ ,
\label{dalc2}
\\
\frac{ {\partial} {\tilde \alpha}_1 }{ {\partial} c_3 } & = &
- \frac{ {\partial} {\tilde \alpha}_2 }{ {\partial} c_3 } = 
\frac{1}{ {\beta}_0^3 ( {\rm det} )^{3/2} } (f_2-f_1^2)^3 \ .
\label{dalc3}
\end{eqnarray}
In addition, we need in (\ref{PMSeq2}) also
${\partial a (p_i^2) }/{\partial (p_i^2)}$ which is
directly obtained from RGE (\ref{aRGE}), and we need as well
${\partial a (p_i^2) }/{\partial c_j}$. The latter
derivatives were derived by Stevenson \cite{Stevenson}
\begin{eqnarray}
\frac{ {\partial} a }{ {\partial} c_2 } & = & 
a^3 \left[ 1 + \frac{1}{3} c_2 a^2 + \left(
- \frac{1}{6} c c_2 + \frac{1}{2} c_3 \right) a^3 +
\cdots \right] \ ,
\label{dac2}
\\
\frac{ {\partial} a }{ {\partial} c_3 } & = & 
\frac{1}{2} a^4 \left[ 1 - \frac{c}{3}a +
\frac{c^2}{6} a^2 + \left(- \frac{c^3}{10}+ \frac{c c_2}{15}
+ \frac{c_3}{5} \right) a^3 + \cdots \right] \ .
\label{dac3}
\end{eqnarray}
Inserting expressions (\ref{duc2})-(\ref{dac3}) and (\ref{aRGE})
into PMS relations (\ref{PMSeq2}), we obtain explicit
equations for the PMS improvement of the dBGA's
for the case of truncated perturbation series $S_n$
of Eq.~(\ref{Spn}) with $n\!=\!3$ (and $\ell\!=\!1$)
\begin{eqnarray}
\frac{ {\partial} }{ {\partial c}_2 } 
\left( a G_f^{[1/2]} \right) &=&
+\frac{3}{ {\beta}_0^2 ( {\rm det} )^{3/2} }
A_2^2 \left[ A_1 - \frac{ 5 \beta_1 }{3 \beta_0^2 } A_2 \right]
\left[ a(p_1^2) - a(p_2^2) \right]
\nonumber\\
&&+ \sum_{i=1}^2
{\tilde \alpha}_i {\Bigg \{} - a^2(p_i^2)
\left[ 1 + c a(p_i^2) + c_2 a^2(p_i^2) + c_3 a^3(p_i^2) +
\cdots \right] \times 
\nonumber\\
&& \times \frac{1}{ \beta_0 A_2} \left[
 - f_1 - \frac{ 5{\beta}_1 }{ 4{\beta}_0^2 } 
- 3 (-1)^i \left( A_2^2 + \frac{5 \beta_1 }{ 12 \beta_0^2 } A_1
\right) \frac{1}{ \sqrt{{\rm det}} } 
  + {\tilde u}_{i}  \right]
\nonumber\\
&& 
+ a^3(p_i^2) \left[ 1 + \frac{c_2}{3} a^2(p_i^2) +
\left( - \frac{ c c_2 }{6} + \frac{ c_3 }{2} \right) a^3(p_i^2)
+ \cdots \right] {\Bigg \}}  = 0 \ ,
\label{PMSc2}
\\
\frac{ {\partial} }{ {\partial c}_3 } 
\left( a G_f^{[1/2]} \right) &=&
+\frac{1}{ {\beta}_0^3 ( {\rm det} )^{3/2} }
A_2^2 \left[ a(p_1^2) - a(p_2^2) \right]
\nonumber\\
&&+ \sum_{i=1}^2
{\tilde \alpha}_i {\Bigg \{} - a^2(p_i^2)
\left[ 1 + c a(p_i^2) + c_2 a^2(p_i^2) + c_3 a^3(p_i^2) +
\cdots \right] \times 
\nonumber\\
&& \times \frac{1}{4 {\beta}_0^2 A_2}
\left[ 1 + (-1)^i \frac{ A_1 }{ \sqrt{{\rm det}} } \right] 
+ \frac{1}{2} a^4(p_i^2) \times
\nonumber\\
&& \times \left[
1 - \frac{c}{3} a(p_i^2) + \frac{c^2}{6} a^2(p_i^2) +
\left( - \frac{c^3}{10} + \frac{c c_2}{15} + \frac{c_3}{5}
\right) a^3(p_i^2) + \cdots \right] {\Bigg \}} = 0 \ ,
\label{PMSc3}
\end{eqnarray}
where we denoted
\begin{equation}
A_1 = (f_3 - 3 f_1 f_2 + 2 f_1^3) \ , \qquad
A_2 = (f_2 - f_1^2) \ .
\label{notA}
\end{equation}
All the parameters $f_j$, ${\rm det}$, ${\tilde \alpha}_i$,
$p_i^2$ appearing in (\ref{PMSc2})-(\ref{notA})
are given explicitly in Ref.~\cite{Cvetic}
in terms of the three TPS coefficients
$r_1,\!r_2,\!r_3$. The latter coefficients, when
known in one RSch, are known in any RSch since their
dependence on $c_j$ ($j\!\geq\!2$) is determined
according to (\ref{r1r2})-(\ref{r3}). We stress that
$a(p_i^2)$ in (\ref{PMSc2})-(\ref{PMSc3}) are
determined by evolving $a(p^2)$ from a chosen
RScl $q^2$ to $p_i^2\!=\!p_i^2(c_2,\!c_3,\!\ldots)$
via RGE (\ref{aRGE}) where the coefficients
$\beta_j\!\equiv\!\beta_0 c_j$ ($j\!\geq\!2$) on the
right are the ones of the RSch used in 
Eqs.~(\ref{PMSc2})-(\ref{PMSc3}). In this evolution,
the initial value $a(q^2;\!c_2,\!c_3,\!\ldots)$
is also known (calculable), once it is known in 
one specific RSch $a(q^2;\!c_2^{(0)},\!c_3^{(0)},\!\ldots)$
-- cf. Eq.~(\ref{RSchconn2}), or (\ref{RSchconn3}).

We note that the obtained system of equations
(\ref{PMSc2})-(\ref{PMSc3}) for PMS improvement 
is relatively complicated
and can be solved only numerically. It results in
finding optimal RSch parameters $c_2,\!c_3$ (i.e., 
${\beta}_2,\!{\beta}_3$) -- optimal in a PMS sense. 
A more limited goal of achieving local independence of
only the RSch-parameter $c_2$ leads us to an easier task
of solving numerically only Eq.~(\ref{PMSc2}), by varying
${\beta}_2$ parameter and keeping ${\beta}_3$ at a fixed value
of a specific scheme.

Incidentally, a very analogous 
kind of explicit PMS improvement 
equations can be constructed also 
for the case of the dBGA approximant
to a TPS $S_n$ of Eq.~(\ref{Spn}) 
with $n\!=\!2$ (and $\ell\!=\!1$),
i.e., for the dBGA's discussed and constructed in the
previous Section. In this case, expressions are simpler. It
is straightforward to check that 
in this case the PMS-improved
expression for the dBGA (\ref{dBGAres}) with respect to
the RSch-parameter $c_2$ is the one satisfying
\begin{eqnarray}
\frac{ {\partial} }{ {\partial c}_2 } 
\left( a G_F^{[1/2]} \right) &=&
\frac{6}{ ( {\tilde b}_1^2 - 4 {\tilde b}_2 )^{3/2} }
\left[ a(p_1^2) - a(p_2^2) \right]
- \frac{3}{ ( {\tilde b}_1^2 - 4 {\tilde b}_2 ) }
{\Big \{} \left[ a^2(p_1^2) + a^2(p_2^2) \right] +
\nonumber\\
&& + c \left[ a^3(p_1^2) + a^3(p_2^2) \right]
+ c_2 \left[ a^4(p_1^2) + a^4(p_2^2) \right] + \cdots
{\Big \}}
\nonumber\\
&& + \frac{1}{ \sqrt{ {\tilde b}_1^2 - 4 {\tilde b}_2 } }
{\Big \{} \left[ a^3(p_1^2) - a^3(p_2^2) \right]
+ \frac{c_2}{3} \left[ a^5(p_1^2) - a^5(p_2^2) \right]
+ \cdots {\Big \}} = 0 \ .
\label{PMSc2m}
\end{eqnarray}

We emphasize that the renormalization scale
$q^2$ in various schemes in this formalism is defined to
have the same meaning, i.e., that the energy parameter
${\tilde \Lambda}$ appearing in the parameter
${\tau}\!=\!{\beta}_0 \ln ( q^2/{\tilde \Lambda}^2 )$
is the same in all schemes under consideration
(cf.~\cite{Celmaster} and Appendix A of \cite{Stevenson}
for details). Stated otherwise, the gauge coupling parameters
in various RSch's behave, with this definition of the
RScl $q^2$, in the following way:
\begin{equation}
a(q^2;\!{\bar c}_2,\!{\bar c}_3,\!\ldots) =
a(q^2;\!c_2,\!c_3,\!\ldots) \left[
1 + {\cal {O}}(a^2) \right] \ .
\label{RSchconn1}
\end{equation}
Specifically, as implied by Eqs.~(\ref{dac2}) and (\ref{dac3})
by Taylor expansion, we have the connections
\begin{eqnarray}
a(q^2;\!{\bar c}_2,\!{\bar c}_3,\!\ldots) &=&
a(q^2;\!c_2,\!c_3,\!\ldots) {\Big [}
1 + ( {\bar c}_2 - c_2 ) a^2(q^2;\!c_2,\!c_3,\!\ldots) +
\nonumber\\
&& + \frac{1}{2} ( {\bar c}_3 - c_3 ) a^3(q^2;\!c_2,\!c_3,\!\ldots)
+ \cdots {\Big ]} \ .
\label{RSchconn2}
\end{eqnarray}
More exactly, the connection between 
$a\!\equiv\!a(q^2;\!c_2,\!c_3,\!\ldots)$ and
${\bar a}\!\equiv\!a(q^2;\!{\bar c}_2,\!{\bar c}_3,\!\ldots)$ 
is given by the following equation 
(Appendix A of Ref.~\cite{Stevenson}):
\begin{eqnarray}
\lefteqn{
\frac{1}{a} + c \ln \left( \frac{ ca }{ 1+ca } \right)
+ \int_0^a dx \left[ - \frac{1}{x^2 (1+cx+c_2 x^2 + c_3 x^3 +
\ldots )} + \frac{1}{ x^2 (1+cx) } \right] = }
\nonumber\\
&& =
\frac{1}{{\bar a}} + c \ln \left( 
\frac{ c{\bar a} }{ 1+c{\bar a} } \right)
+ \int_0^{{\bar a}} dx \left[ - 
\frac{1}{x^2 (1+cx+{\bar c}_2 x^2 + {\bar c}_3 x^3 +
\ldots )} + \frac{1}{ x^2 (1+cx) } \right] \ ,
\label{RSchconn3}
\end{eqnarray}
which can be solved numerically for ${\bar a}$.
Eq.~(\ref{RSchconn2}), or (\ref{RSchconn3}),
determines then the initial value
$a(q^2)$ for integration of RGE (\ref{aRGE}) 
from $p^2\!=\!q^2$ to $p^2\!=\!p_i^2$ in
{\em any\/} RSch $({\bar c}_2,\!{\bar c}_3,\!\ldots)$,
once it is known in one specific RSch $(c_2,\!c_3,\!\ldots)$.

For example, in the specific case of QED with the on-shell (OS) and the
${\overline {\rm MS}}$ schemes, the above relation 
(\ref{RSchconn2}) gives
us immediately the connection between the fine structure constant 
${\alpha}_{\rm f.s.}/{\pi}\!\equiv\!a_{\rm f.s.}\!\equiv\!a(q^2\!=\!-m_e^2
)^{\rm on-shell}$
and the ${\overline {\rm MS}}$ constant ${\bar a}(q^2\!=\!-m_e^2)$
\begin{equation}
{\bar a}(q^2\!=\!-m_e^2) = a_{\rm f.s.} \left[ 1 +
0.9375 a^2_{\rm f.s.} + (0.07131285\!\ldots) a^3_{\rm f.s.}
+ \cdots \right] \ ,
\label{MSOS}
\end{equation}
which agrees with the result of \cite{Broadhurst}. 
In order to obtain (\ref{MSOS}), it was enough to insert into relation
(\ref{RSchconn2}) the known QED beta coefficients 
${\beta}_j\!=\!{\beta}_0 c_j\!=\!- c_j/3$ in the two schemes
\cite{QEDbeta}, in the convention of Eq.~(\ref{aRGE})
\begin{eqnarray}
{\beta}_2({\rm OS})^{\rm (QED)} & = & + 0.420138888\!\ldots \ ,
\quad
{\beta}_3({\rm OS})^{\rm (QED)}   =   + 0.571156328087\!\ldots \ ,
\nonumber\\
{\beta}_2({\overline {\rm MS}})^{\rm (QED)} 
& = & + 0.107638888\!\ldots \ ,
\quad
{\beta}_3({\overline {\rm MS}})^{\rm (QED)} 
 =  + 0.523614426964\!\ldots \ ,
\label{betaOSMS}
\end{eqnarray}
while ${\beta}_0^{\rm (QED)}\!=\!-1/3$ and 
${\beta}_1^{\rm (QED)}\!=\!-1/4$ are RSch-invariant.
We mention these relations in order to stress that
similar relations (\ref{RSchconn1})-(\ref{MSOS})
between the on-shell and the (nonmodified) MS schemes
are not true, because the meaning of the RScl in MS scheme
is different from that of the ${\overline {\rm MS}}$
scheme (and hence of the OS scheme) 
by a constant factor. In a way, MS and
${\overline {\rm MS}}$ schemes can be regarded as the
same, except that the meaning of the RScl $q^2$ differs
in them by a constant factor. Therefore, although 
conditions (\ref{PMSc2})-(\ref{PMSc3}) and (\ref{PMSc2m})
search for a PMS-improved dBGA approximant only in a
certain class of renormalization schemes (those whose
RScl $q^2$ has the same meaning), these conditions nonetheless
don't ``miss'' any relevant scheme.

For practical purposes, we should keep in mind that
a given TPS $S_n$ [cf. Eq.~(\ref{Spn})] is always given in
such a specific RSch (e.g., on-shell or MS) in which only
the first four coefficients ${\beta}_j$ ($j\!=\!0,\!1,\!2,\!3$)
on the right of RGE (\ref{aRGE}) are known. This means
that the changing of $a(q^2;\!c_2,\!c_3,\!\ldots)$ when
RSch and RScl are changed is practically known only up to
(and including) $\sim\!a^4$, as seen, e.g., from (\ref{RSchconn2})
and Eq.~(54) of Ref.~\cite{Cvetic} (first entry).
This, together with relations (\ref{r1r2})-(\ref{r3}), 
implies that we can consistently define the TPS
$S_n$ in various RSch's only when $n\!\leq\!3$, i.e.,
when only at most three coefficients $r_j$ ($j\!=\!1,\!2,\!3$)
are available. Therefore, the described PMS-improved method of
modified dBGA's can at present be consistent in practice only
for such TPS's (this is the case also in the usual PMS
approach). Incidentally, the available TPS's $S_n$ at
present have $n\!\leq\!3$ (in QCD: $n\!\leq\!2$).

Within this context, the general limitations on the available 
precision of the value of a given coupling parameter 
$a(q^2;\!c_2,\!\ldots)$, and hence of the values of the obtained 
approximants, become apparent. 
In QCD (and QED), only the first four ${\beta}_j$
coefficients ($j\!=\!0,\!\ldots,\!3$) are known in a given
specific (``original'') RSch (in QCD: MS or ${\rm {\overline {MS}}}$; 
in QED: on-shell) in which
the value of that coupling parameter is fairly well known
at a certain original RScl $q^2$. We can then calculate the coupling 
parameter $a(q^{\prime 2};\!{\bar c}_2,\!\ldots)$ in any other RSch
and at any other RScl only up to (and including) $\sim\!a^4$. 
Therefore, even if we knew the value of $a(q^2;\!c_2,\!\ldots)$ with
very high precision in the original RSch and at the original
RScl, the values of $a$ in other RSch's and at other
RScl's would be available only within the limited precision
$\sim\!a^4$. However, the PMS-improved modified
dBGA approximants would not be so crucially affected by
this limited precision of the values of $a$ in various
RSch's and at various RScl's. The reason lies in the fact that
these approximants (as well as the usual PMS approximants),
besides being dependent on the value of 
$a(q^{\prime 2};\!{\bar c}_2,\!\ldots)$,
depend solely on the first $n$ ($\leq\!3$) coefficients 
$r_j(q^{\prime 2};\!{\bar c_2},\!\ldots,\!{\bar c}_j)$
($j\!=\!1,\!\ldots,\!n$) in the TPS $S_n$, and these coefficients
are RScl- and RSch-dependent only via the values of the
RScl $q^{\prime 2}$ [i.e., 
${\tau}^{\prime}\!=\!\ln(q^{\prime 2}/{\tilde {\Lambda}}^2)$]
and of the first $n\!-\!1$ RSch-parameters 
${\bar c}_j\!=\!{\beta}_0 {\bar {\beta}}_j$ ($j=\!2,\ldots,\!n$),
as seen explicitly in (\ref{r1r2})-(\ref{r3}).

\section{Conclusions}

Based on the previously introduced method of modified diagonal
Baker-Gammel approximants (dBGA's) \cite{Cvetic} for a specific
class of truncated perturbation series (TPS's) of observables, 
we showed how to extend the applicability
of the method to {\em any\/} TPS of any given observable.
The approximants reproduce the TPS, to which they are
applied, to the available order of precision, and they
are exactly invariant under the change of the
renormalization scale (RScl). Furthermore, we constructed
equations for these dBGA's which have to be satisfied
in order to have minimal (zero) sensitivity to the local 
change of the renormalization scheme (RSch) in these
dBGA's. The latter conditions are
just a variant of the method of the principle of the minimal
sensitivity (PMS) with respect to changing the RSch-parameters
${\beta}_j$ ($j\!\geq\!2$), but this time applied to the
obtained dBGA's and not to the TPS's. The resulting approximants 
(dBGA's) are then RScl- {\em and\/} RSch-invariant, i.e., independent
of the RScl $q^2$ and the RSch ${\beta}_j$ ($j\!\geq\!2$)
chosen in the TPS under consideration.

The described dBGA method is an improvement of the method of the
diagonal Pad\'e approximants (dPA), the latter being
RScl-invariant only in the large-$|{\beta}_0|$
limit and RSch-noninvariant. We believe that there is
some room for optimism concerning the efficiency of
the described dBGA method when compared with the usual PMS method
\cite{Stevenson}, and with the ECH method \cite{Grunberg}
which in turn is related with the usual PMS.
This optimism rests on the fact that the
RScl-invariance in the described dBGA approach is
ensured via an algorithm which keeps a close contact with
the usual (d)PA method, and the latter method has proven
to be reasonably efficient for various QCD observables
\cite{PAQCD}. The RSch-invariance of the described
dBGA's, i.e., invariance under the change of the 
${\beta}_j$'s ($j\!\geq\!2$) appearing in the original TPS's,
is achieved in a way similar to the usual PMS approach.
The latter approach, on the other hand, achieves the
RSch- and RScl-invariance by requiring that the
truncated perturbation series (TPS) itself have
minimal (zero) sensitivity under the local change of the
RScl ($q^2$) and the RSch (${\beta}_j$'s, $j\!\geq\!2$).

In order to test the quality of the described
method of the PMS-improved modified dBGA's in 
practical calculations, it would be necessary to compare
results of this method with those of the 
PMS \cite{Stevenson} and ECH \cite{Grunberg} method
-- in the cases of various available
truncated perturbation series of QCD and QED observables.
Further, comparison with several other methods 
\cite{Maxwell}-\cite{SolovtsovShirkov} which
eliminate or reduce the RScl- and RSch-dependence
would give additional insights into the question
of the value of the presented approximants.

\vspace{1.cm}

\noindent {\large {\bf Acknowledgments:}}

\noindent One of the authors (G.C.)
wishes to thank Professor D.~Schildknecht for offering
him financial support of Bielefeld University during part of
this work. He also wishes to thank M.~Kalmykov
for valuable discussions concerning renormalization schemes.
He further acknowledges the hospitality of DESY Hamburg
and MPI Munich, where part of this work was done.

\vspace{1.cm}

\noindent 
{\footnotesize
{\bf Abbreviations used frequently in the 
article\/:}
(d)BGA -- (diagonal) Baker-Gammel approximant; 
(d)PA -- (diagonal) Pad\'e approximant; 
ECH -- effective charge (method);
PMS -- principle of minimal sensitivity;
RSch -- renormalization scheme; 
RScl -- renormalization scale;
TPS -- truncated perturbation series.}

\end{document}